# RapidStream IR: Infrastructure for FPGA High-Level Physical Synthesis

Jason Lau[†1,2], Yuanlong Xiao[†1], Yutong Xie[1], Yuze Chi[1], Linghao Song[2],
Shaojie Xiang[3], Michael Lo[2], Zhiru Zhang[1,3], Jason Cong[1,2], Licheng Guo[1]

[1]RapidStream Design Automation, Inc.   [2]University of California, Los Angeles   [3]Cornell University

{lau,lcguo}@rapidstream-da.com

## ABSTRACT

The increasing complexity of large-scale FPGA accelerators poses significant challenges in achieving high performance while maintaining design productivity. High-level synthesis (HLS) has been adopted as a solution, but the mismatch between the high-level description and the physical layout often leads to suboptimal operating frequency. Although existing proposals for high-level physical synthesis, which use coarse-grained design partitioning, floorplanning, and pipelining to improve frequency, have gained traction, they lack a framework enabling (1) pipelining of real-world designs at arbitrary hierarchical levels, (2) integration of HLS blocks, vendor IPs, and handcrafted RTL designs, (3) portability to emerging new target FPGA devices, and (4) extensibility for the easy implementation of new design optimization tools.

We present RapidStream IR, a practical high-level physical synthesis (HLPS) infrastructure for representing the composition of complex FPGA designs and exploring physical optimizations. Our approach introduces a flexible intermediate representation (IR) that captures interconnection protocols at arbitrary hierarchical levels, coarse-grained pipelining, and spatial information, enabling the creation of reusable passes for design frequency optimizations. RapidStream IR improves the frequency of a broad set of mixed-source designs by 7% to 62%, including large language models and genomics accelerators, and is portable to user-customizable new FPGA platforms. We further demonstrate its extensibility through case studies, showcasing the ability to facilitate future research.

## 1 INTRODUCTION

The evolution of FPGAs into larger, multi-die devices, e.g., the two-die Versal VHK158 and the three-die Alveo U280, has improved their ability to accelerate complex computations, such as large language models, enhancing performance and energy efficiency [8, 45]. As FPGA designs become more sophisticated, the designers increasingly rely on high-level synthesis (HLS) to manage complexity, which allows them to describe designs at the algorithmic level and generate RTL code, thereby reducing development effort [10].

HLS's abstraction comes with its own set of challenges, primarily concerning the physical optimization. The absence of cycle-accurate and physical layout information in the untimed HLS specification often leads to a mismatch between the frontend HLS and the back-end physical implementation, hindering timing closure [17, 18]. The scalability issues with current EDA tools exacerbate this problem, where duplicating processing logic without manual floorplanning can adversely affect the quality of results (QoR) [27, 40].

Compared to the HLS approach, an RTL expert, cognizant of a module's resource-intensive feature and potential for substantial area occupation, might manually add adequate pipeline levels between neighboring modules to break the delay of long wires crossing dies. Most HLS tools, however, lacking such architectural information, often fail to apply this optimization, leading to long critical paths. Moreover, an RTL expert could divide the design into groups and assign them to specific dies to balance resource utilization. Unfortunately, for HLS tools, there are usually insufficient pipelining levels between generated logic blocks. This forces downstream tools to place these blocks closer together to minimize total wire length, which in turn causes local routing congestion.

In response, researchers have proposed techniques that we define as **high-level physical synthesis (HLPS)**, integrating coarse-grained design partitioning, floorplanning, and pipelining to co-optimize HLS with physical design stages, aiming for improved frequency [12, 16, 17, 25, 30, 32, 33, 42, 43]. With HLPS, the design is partitioned into coarse-grained groups, which are then roughly floorplanned on the FPGA to minimize the number of connections between dies. Based on the layout information, pipeline registers are added to break up long connections between the groups, allowing the communication to operate over multiple clock cycles instead of one cycle with long latency. Additionally, the resource requirements of each design module are analyzed to be balanced across the FPGA to avoid localized congestion and mitigate the critical paths caused by limited routing resources. One representative work using this methodology is the AutoBridge framework [17], which helps create high-frequency FPGA HLS designs that span multiple dies by reducing the impact of long connections between dies and evenly distributing the design across the available resources.

However, existing proposals for FPGA HLPS have several limitations that make them difficult to apply to real-world designs.

For instance, these research works often focus on a limited set of designs generated by AMD/Xilinx Vitis HLS, and:

**(1) They do not support design optimizations at arbitrary hierarchical levels**; therefore, all task-level parallel modules need to be interconnected at the top HLS function [17, 42].

**(2) They cannot integrate handcrafted RTL and vendor IPs**, despite real-world HLS designs often including these components.

**(3) They are limited to specific FPGA devices**, such as the AMD Alveo U250 and U280, making it challenging to adapt them to hardware that satisfies specific compute or budget requirements.

**(4) They lack an extensible infrastructure** for exploring different research directions in partitioning and pipeline schemes.

We propose RapidStream IR (RIR), a composition and exploration infrastructure for FPGA HLPS. It supports hybrid-source FPGA HLS designs and customizable FPGA devices, optimizing them to achieve high frequency. This infrastructure tackles the aforementioned technical challenges with the following key features:

**Intermediate representation (IR)** - Our solution offers a flexible and extensible IR of the input design that can be transformed using any programming language. This IR effectively captures the connectivity, the ability to pipeline components throughout the hierarchical structure, and the spatial information of the design. RIR ensures that the functionality of the design remains intact throughout transformations or manipulations performed on the IR.

**Reusable design optimization passes** - RIR provides a set of reusable passes for transforming the design, such as hierarchical rebuilding, module partitioning, and module insertion. By leveraging these passes, researchers can explore different optimization strategies and easily tailor the framework for specific design targets.

**Support for diverse formats of designs** - Analyzers for different design formats are provided, such as Verilog, Xilinx Compiled IPs (XCI), and Vitis HLS-generated designs. The framework is extensible to other source formats, such as Dynamatic HLS [22, 23] and Catapult HLS [37], by implementing information extractors.

**Portability to various platforms** - RIR ensures portability across different FPGA platforms by offering an interface to define new devices without altering the analyzers or optimization passes.

Consider the large language model (LLM) FPGA accelerator [8] in Figure 1 as a motivating example. This design incorporates various source formats, including HLS, RTL, and Xilinx IP blocks, linked hierarchically using Verilog. Specifically, the data loaders and buffers are implemented using hand-written RTL, while the computational kernels are generated through HLS. The entire architecture is interfaced with external memory via Xilinx IP blocks.

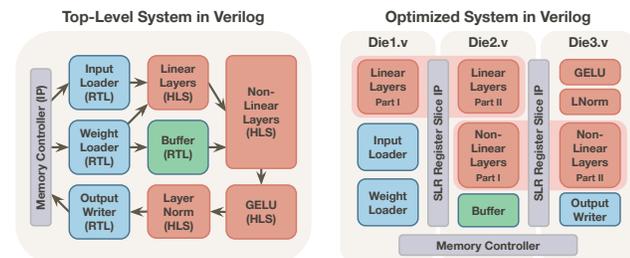

**Figure 1: The FPGA HLS accelerator design for large language models (LLM) [8] before and after physical optimizations.**

The initial design operates at 150 MHz on the Xilinx Alveo U280 FPGA. Chen *et al.* [8] improve this by manually distributing modules among FPGA dies and adding registers to break up critical paths, as shown in Figure 1, raising the frequency to 245 MHz. This approach involves partitioning modules, such as linear layers, across multiple FPGA regions, which complicates code management. The lack of an HLPS framework that supports mixed-source integration poses challenges in automating this time-consuming and error-prone process. Moreover, adapting the design for new or customized hardware requires substantial changes to the RTL hierarchy and layout constraints for optimal performance.

RIR automates the optimization on the LLM FPGA design by composing multiple design source formats and exploring partitioning and pipelining strategies. Using RIR, a comparable frequency of 243 MHz is achieved on the U280 FPGA without code modifications and can be ported seamlessly to other FPGAs. Evaluations on six FPGA devices show frequency improvements ranging from 30% to 62%, maintaining an average frequency of 244 MHz.

Our technical contributions are as follows:

(1) We present RapidStream IR (RIR), the first HLPS infrastructure that supports the hierarchical composition of FPGA designs from diverse sources, such as HLS-generated modules, RTL, and vendor IPs. This framework enables the exploration of physical optimizations in complex FPGA designs, aiming to achieve high frequency while maintaining design productivity.

(2) We introduce a flexible and extensible IR for HLPS. This approach allows for the creation of reusable passes that cater to various design formats and device targets, requiring only the implementation of minimal information extractors.

(3) Through case studies, including floorplan exploration, parallel synthesis, and design debugging, we demonstrate the extensibility of our framework. These studies highlight RIR's ability to facilitate research and exploration, and they suggest future enhancements and applications of the framework.

RIR enables the exploration of global timing optimization for hierarchically-composed HLS designs and adapts to new platforms such as Versal VP1552. Experimental results show a cutting-edge frequency improvement of 40% on average across well-researched and newly introduced FPGA platforms.

## 2 BACKGROUND
### 2.1 Challenges in Large-Scale FPGA Designs

FPGA HLS design optimization poses challenges due to architectural variations in resource distribution and wire latency across and even within devices. As a result, designers encounter difficulties with timing optimization and design portability. Figure 2 illustrates this issue using three representative device examples [1, 17]:

(1) The AMD Alveo U55C FPGA has three dies, each with some resources dedicated to the Vitis shell. It connects to 32 High-Bandwidth Memory (HBM) channels at the bottom, with unprogrammable gap regions in the center. In Vitis HLS, HBM channels are accessed via pointers, with no control over accessor module's location or pipeline levels to the HBM controller.

(2) The AMD Versal VP1552 FPGA consists of two dies with six and seven regions in height, respectively. It features networks-on-chip and an integrated ARM processor. Discontinuities caused by the IPs may extend routing paths for nearby signals, and die crossings incur notably higher latency, unreflected in HLS.

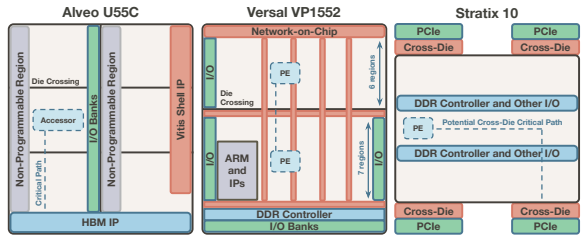

Figure 2: Layout of FPGA devices. Die boundaries incur substantial latency, and the gap regions and IPs limit resource utilization. These limitations are not considered in HLS.

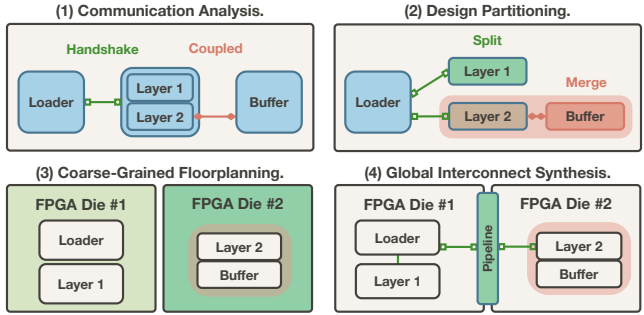

Figure 3: HLPS for the first three stages of the LLM design [8].

(3) The Intel Stratix 10 FPGA has I/O banks at the center of the programmable logic, with multi-die interconnect bridges and PCIe blocks on the sides, which are not modeled in HLS.

## 2.2 High-Level Physical Synthesis (HLPS)

*High-level physical synthesis (HLPS)* is proposed to bridge the gap between HLS and physical design. By providing HLS with the FPGA's physical layout, it can effectively partition and floorplan modules, and identify long wires for pipeline stage insertion. Figure 3 illustrates the HLPS flow using the first few stages of the LLM accelerator. The process can be summarized in the following stages:

(1) **Communication Analysis.** The high-level specification, such as C++ code, is analyzed to identify connections between module units that can tolerate latency, such as handshakes, "data-valid" protocols, and interconnect buses. These communications are typically represented as streams, data flow regions, global pointers, and function arguments in C++.

(2) **Design Partitioning.** The design is divided into partitions based on communication patterns, allowing only latency-tolerant connections between groups. These partitions can be distributed across distant regions or different dies, and the connections between them can be pipelined, breaking global critical paths.

(3) **Coarse-Grained Floorplanning.** Partitions are allocated to coarse-grained regions on the FPGA, optimizing multiple objectives such as minimizing inter-region wire crossings, managing regions with limited available resources, and balancing resource distribution to prevent local routing congestion.

(4) **Global Interconnect Synthesis.** Once the location of each partition is determined, the partitions are interconnected based on estimated delay to break critical paths for timing closure.

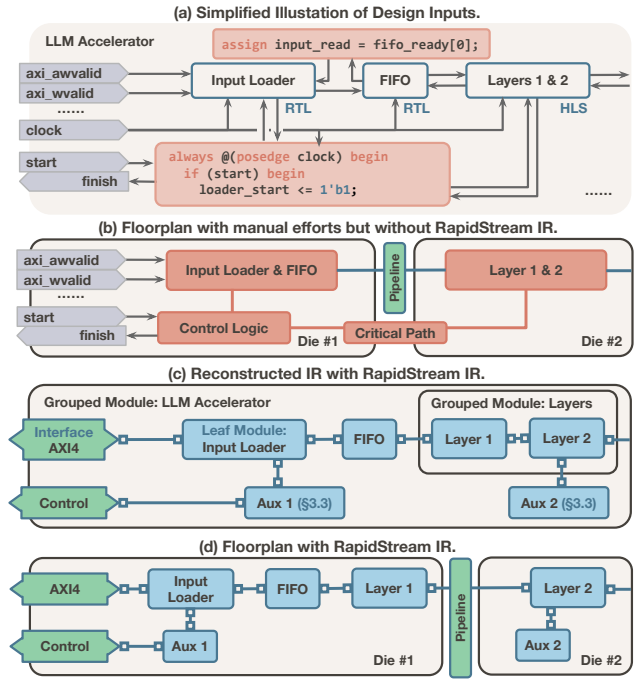

Figure 4: The first three modules of the LLM accelerator [8] and the optimizations with or without RapidStream IR.

A number of studies [12, 16, 17, 25, 30, 32, 33] have investigated methodologies for HLPS and demonstrated their effectiveness in automatically optimizing the frequency of HLS designs.

## 2.3 Motivating Example

The application of HLPS is limited by several shortcomings in (1) global optimization across hierarchical levels, (2) integration of diverse source formats, and (3) adaptation to new devices. These infrastructural shortcomings not only limit the practical deployment of HLPS but also impede ongoing research, as researchers must develop non-reusable custom tools for each new design or device.

The need for an infrastructure for HLPS can be exemplified by the LLM FPGA design [8] shown in Figure 4a. This figure shows a simplified segment of the LLM accelerator discussed in Section 1, which interconnects modules using Verilog to integrate RTL, HLS, and IP components. This example is referred to throughout the paper to highlight challenges and discuss our proposed solutions.

In Figure 4a, this LLM accelerator consists of three modules: (1) an Input Loader in Verilog, (2) a Verilog FIFO buffering data, and (3) a hierarchical HLS kernel consisting of the initial two linear LLM layers. These two Layers have a collective role and are organized as a single HLS function, which invokes two subfunctions to perform its operations. The top-level interconnect is in Verilog, which not only serves to instantiate each submodule but also incorporates control logic using `assign` and `always` statements.

Existing HLPS methods are inadequate for optimizing this LLM design due to their lack of support for RTL components such as the Input Loader and the FIFO. Additionally, they fail to interpret the top-level Verilog logic, hindering communication analysis, such as those linked by `assign` statements. Even if RTL support were present, the outcome would remain suboptimal. As depicted in Figure 4b, the absence of hierarchical pipelining forces resource-intensive Layers 1 and 2 to be placed together on the same die, leading to local congestion. Furthermore, without the ability to understand Verilog code, the top-level control logic is treated as a monolithic module, resulting in non-pipelined connections to the Layers that contribute to the creation of global critical paths.

RapidStream IR tackles these challenges by first converting the design into an equivalent intermediate representation (IR), as depicted in Figure 4c. This IR is then optimized through a series of transformation passes into the optimized form shown in Figure 4d. RapidStream IR enhances HLPS with: **(1) hierarchical optimization**, which allows the independent placement and pipelining of Layers 1 and 2 to balance resource requirements; **(2) format-agnostic analysis**, enabling the integration of the Input Loader and the FIFO, and allowing the partitioning of Verilog control logic into two independent units, Aux 1 and 2 (to be discussed in §3.1); and **(3) portability across input formats and FPGA platforms** by unifying information in the IR, such as interface and pipelinability.

## 3 RAPIDSTREAM IR FRAMEWORK

RapidStream IR (RIR) consists of three components: (1) a progressively refined **intermediate representation (IR)** that remains agnostic to specific HLS frameworks, EDA tools, or coding styles; (2) **utility plugins** that facilitate the input of design specifications and output to EDA tools; and (3) a set of reusable **transformation passes** for composing design optimizations. Each element plays a vital role in enhancing the HLPS flow. In the absence of an IR, HLPS researchers are forced to analyze the Verilog code for module communication and insert pipelines directly into the code. Lacking utility plugins, researchers must manually engage EDA tools for resource analysis and for setting design constraints.

**Big Picture.** The overall architecture of RIR is shown in Figure 5. It takes three *inputs*: FPGA design (e.g., Verilog, netlists), their high-level interface information (e.g., HLS reports, pragmas), and Python directives for device information and EDA tool interaction. These inputs are processed by *plugins* into the *IR*. Transformation *passes* then modify the IR to perform the HLPS flow. The final IR is processed by *plugins* back into design code and layout hints for EDA tool implementation. Users can also write tools to modify the output IR for customization. Integrating them, we implement a complete HLPS system that optimizes real-world designs in Section 3.4.

**Design Principles.** We design with the following in mind:
(1) **Enabling Incremental Analysis and Transformation.** We intentionally limit the IR to be lightweight and robust. By providing a set of composable core passes that reduce a complex design into our canonical form and progressively obtain the required information, we make the passes and plugins simple.
(2) **Scoping Flexibility.** We focus on the practical requirements for HLPS methodologies, rather than creating an all-in-one solution like MLIR [26]. This approach allows us to prioritize coarse-grained module interactions while maintaining support for fine-grained logic that cannot be easily translated into IR, such as in Xilinx vendor IPs [3] and design netlists.
(3) **No Language "Lock-In".** Not all computations warrant the development overhead of C++. To support all major languages, we make the IR as simple as possible as a subset of the JSON schema [24] and provide automated language binding generators. In fact, we wrote many passes in Rust, Python, and Java, and a visualization and debugging tool in TypeScript.

Additionally, we provide "Design Rule Checking (DRC)" passes to ensure the consistency in design information. We further maintain a mapping between the components of the original design and their transformed counterparts throughout the optimization process, enabling human readability and debuggability.

### 3.1 Progressively Refined IR

RIR is an IR that incrementally incorporates a design's coarse-grained information. Each pass progressively infers the design's hierarchy, connections, and port interface properties. It keeps the original fine-grained logic intact if it is unused in the passes.

**Design Elements.** RIR captures the following of a design:

(1) **Module.** A design entity classified into *grouped module* and *leaf module*. Each module is identified by a name and consists of multiple ports, each having direction and width attributes, interconnecting with other modules. They can incorporate *interface* that identifies the potential pipeline methods of the ports.
(2) **Leaf Module.** A basic design unit treated atomically by HLPS, which keeps it intact. Leaf modules can be in any format, such as RTL or IPs, provided they are supported by subsequent EDA tools. RIR provides various utility plugins to obtain the required attributes of a leaf module. A leaf module may be progressively reconstructed into a grouped module or partitioned into multiple leaf modules using RIR's transformation passes.
(3) **Grouped Module.** A reconstructed hierarchy from a leaf module, organizing submodules. Grouped modules act only as containers without adding logic, which implies that each submodule connection must be via a single identifier. RIR passes progressively partitions its submodules while adhering to this rule.
(4) **Interface.** A pipeline strategy that can be applied to a set of ports. The *type* of the interface guides the pipelining strategy, such as handshake or feedforward. When a port is included in an interface, it allows for pipelining by introducing additional pipeline stages. For instance, a *feedforward* interface, carrying only scalar signals, can be pipelined by inserting a flip-flop to break critical paths. A *handshake* interface, involving valid, ready, and data ports, can be pipelined by adding a relay station [6] or an almost-full FIFO [18]. Figure 6 illustrates these two most common interfaces and their pipelining methods.
(5) **Additional Metadata.** The IR can include extra data such as floorplan constraints, resource utilization, and timing characteristics, appended to any IR node as additional fields and progressively inferred and updated by analysis passes as needed.

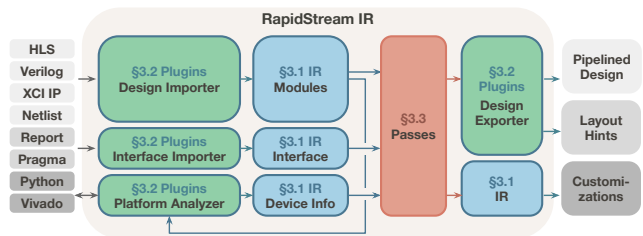

Figure 5: RapidStream IR's overall architecture, consisting of the IR (blue, §3.1), plugins (green, §3.2), and passes (red, §3.3).

**Invariant Assumptions.** During transformation, a few assumptions about the IR are maintained: (1) Each wire in a grouped module must connect precisely two modules, prohibiting fan-out. (2) Each submodule port in a grouped module must connect to only one identifier or a constant, without operations such as concatenation or bit selection. (3) All non-constant ports on an interface should be fully connected to another module, disallowing the splitting or omission of signals. These restrictions maintain the simplicity and ease of manipulation of the IR. Despite being restricted, our core passes enable the transformation of complex designs into this form.

**Virtual Device Definition.** RIR supports a wide range of FPGA devices through virtual device descriptions stored in the IR, which contain the resource distribution within the device and the number of inter-die wires. The virtual device description divides the physical FPGA device into slots. During floorplanning, design modules are mapped to these slots. RIR includes predefined virtual devices for UltraScale+ and Versal, based on empirical data. Users can also customize the virtual device by specifying parameters such as the FPGA device part number and the slot shapes. RIR then uses vendor tools to extract the necessary resource information and automatically generates the virtual device description. Figure 7 shows a virtual device described using our Python API for the Versal VP1552, which partitions the device into two columns and four rows, each containing one-fourth of an FPGA die, by specifying floorplanning rectangles called pblocks in Vivado.

**Sample IR Format.** RIR is structured in formats that the JSON Schema [24] can validate, including data types of dictionaries, lists, strings, and numbers. The choice of storage and exchange format for the IR, such as YAML [5], JSON [7], or XML [38], can optionally vary depending on the programming languages utilized. Figure 8 illustrates a segment of the IR for the LLM accelerator discussed in Section 2.3, presented in YAML format for clarity, alongside its corresponding block graph. The top-level grouped module, LLM (Lines 1-12), instantiates three submodules: InputLoader, FIFO, and Layers (Lines 5-12), which are interconnected via handshake interfaces. InputLoader retrieves text input from memory, FIFO buffers this data, and Layers executes linear layer computations on the buffered input. The IR captures coarse-grained information such as module names (Lines 1, 14), ports (Lines 2-3, 15-19), and wires (Line 4). The instantiation of the FIFO module is denoted as FIFO_inst (Lines 9-11), connecting I_wire to its I port (Line 11). Within the leaf module FIFO, the IR preserves its native form, such as Verilog source code (Line 20). Details regarding the pipeline are specified in the interface section (Lines 21-24), which defines the handshake interface and its associated ports. Each object can

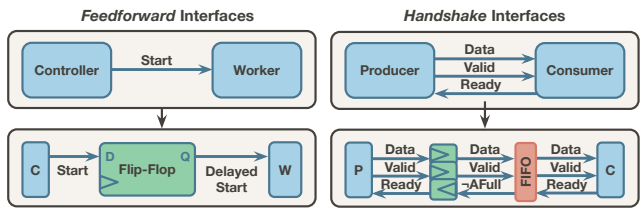

**Figure 6:** *Feedforward* interfaces are pipelined using flip-flop registers, and *handshake* interfaces are pipelined with an almost-full FIFO and registers. AFull indicates that the FIFO is almost full, preventing overflow due to flip-flop latency.

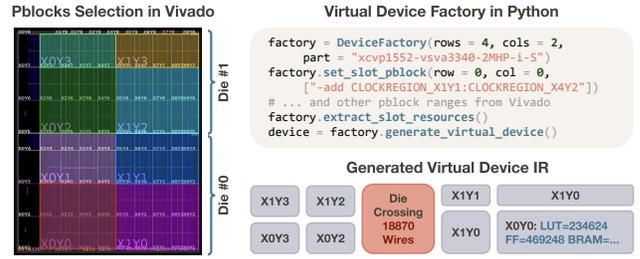

**Figure 7:** Pblocks for VP1552; Virtual device description in Python; Its inferred resource and die-crossing wire capacity.

```yaml
1  - module_name:     LLM
2    module_ports:
3    - { name: ap_clk, direction:  in, width:      1 } # ..
4    module_wires:    [{ name: I_wire, width:     64 }, ..]
5    module_submodules:
6    - instance_name: InputLoader_inst
7      module_name:   InputLoader
8      connections:   [{ port:     I, value: I_wire }, ..]
9    - instance_name: FIFO_inst
10     module_name:   FIFO
11     connections:   [{ port:     I, value: I_wire }, ..]
12   - instance_name: Layers_inst                      # ..
13
14 - module_name:     FIFO
15   module_ports:
16   - { name: I,      direction:  in, width:     64 }
17   - { name: I_rdy,  direction: out, width:      1 }
18   - { name: I_vld,  direction:  in, width:      1 }
19   - { name: ap_clk, direction:  in, width:      1 } # ..
20   module_verilog:  "module FIFO (I, ..); ..; endmodule"
21   module_interfaces:
22   - iface_type:    handshake
23     iface_ports:   { data:      [ I ], clk:  ap_clk ,
24                      ready:   I_rdy, valid:  I_vld }
25   module_metadata:
26     resource:      { FF: 10, LUT: 39, DSP: 0, BRAM: 0,  ..}
27     floorplan:     "SLOT_X1Y1"
```

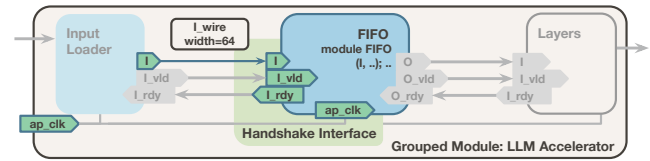

**Figure 8:** Part of the LLM's IR and corresponding block graph.

optionally contain additional metadata specific to different transformation pass, such as resource utilization and floorplan constraints (Lines 25-27). The device information can be embedded in the IR to facilitate transformation passes that optimize the design for a specific FPGA target or to generate constraints for EDA tools.

**Comparisons.** RIR focuses on HLPS for *existing* designs in various formats, unlike the Xilinx IP Integrator (IPI), which assembles IPs into systems. RIR offers a flexible representation that supports incremental analysis through passes. For example, all modules are initially treated as indivisible leaf modules and partitioned as needed. As a result, large HLS-generated modules can be partitioned in RIR, while IPI treats them monolithically.

Accelerator description languages such as Chisel [4] and Calyx [34] enable the high-level specification of fine-grained hardware designs. These languages are orthogonal to RIR, which is tailored for capturing coarse-grained information pertinent to HLPS, including module hierarchy, interfaces, and resource utilization metrics. Given the language-agnostic design of RIR, it can directly incorporate these representations as leaf modules, allowing the transformation of these modules using reusable passes.

On the other hand, MLIR [26] serves as a general-purpose IR across abstraction levels, whereas RIR is tailored for HLPS with a coarse-grained focus. RIR accommodates arbitrary formats in leaf modules, in contrast to MLIR, which mandates details at every level. Although RIR can be represented in MLIR as a custom dialect, such a representation would require C++ for the passes, which RIR intentionally avoids. Moreover, MLIR lacks reusable passes for HLPS, reducing the motivation for its use in this domain.

### 3.2 Utility Plugins

RIR has a suite of utility plugins to bridge the abstract IR and concrete implementations. These plugins are inherently modular, supporting additional source formats and EDA tools as needed. The plugins are categorized into *importers*, *analyzers*, and *exporters*.

**Leaf Module Importer.** It extracts metadata from a module's source format to build a corresponding leaf module in the IR. Parsed data include the module name, ports, etc. To maintain the design integrity, the source code or its binary is directly embedded in the IR. We have developed importers for formats such as Verilog, VHDL, netlists, and Xilinx Compiled IP (XCI). For example, for Verilog, we use Slang [35] to extract module information from the syntax tree. Other formats, such as VHDL, are handled using appropriate parsers or by transforming module signatures into a Verilog stub file using EDA tools, followed by the Verilog importer.

**Interface Importer.** High-level interface information essential for HLPS can be extracted from various sources. Vitis HLS provides interface information in report files, while Xilinx IPs include interface details in XCI files. If interface data is missing, users can provide it using *pragmas in source-code comments* or *interface rules* specified in our Python API with regular expressions. Figure 9 shows a Verilog source code example in which a single-line pragma on Line 5 sets the handshake interface for all 37 AXI ports of the handcrafted memory input loader RTL module.

**Platform Analyzer.** Design optimizations require information from downstream vendor tools, such as resource utilization per module, to balance resource allocation across device regions. The platform analyzer interfaces with vendor tools to collect data.

**Design Exporter.** The design exporter generates the final design output from the IR to be compatible with downstream EDA tools. For unchanged leaf modules, the exporter outputs the original source intact. For modified modules, it generates the corresponding Verilog files. If the IR includes extra metadata, such as floorplanning guidance, the exporter also outputs this data as constraint files.

### 3.3 Composable Transformation Passes

RIR provides a set of transformation passes that progressively gather data and refine the IR to optimize the design. Each pass is designed to "do one thing and do it well," focusing on one aspect to ensure robustness and maintainability, and allowing for easy extension to support new design formats and EDA tools.

In this section, we present the core passes of RIR applied to a subset of the LLM accelerator example [8] to illustrate their functionalities. We use the IR's block graph in Figure 10 for clarity. All modules in the design are imported as leaf modules initially.

**Hierarchy Rebuild Pass.** The rebuild pass converts imported leaf modules into grouped modules to reconstruct the design hierarchy. It creates a grouped module comprising the extracted submodules from the leaf module and its residual logic, which is defined as an **aux (auxiliary) module**. The grouped module has the same

```
1  module InputLoader (
2    output wire m_axi_AWVALID,  input wire m_axi_AWREADY,
3    output wire m_axi_WVALID,   input wire m_axi_WREADY,
4    // ... 33 other AXI ports
5  );
6    // pragma handshake pattern=m_axi_{bundle}{role} \
7    //     role.valid=VALID role.ready=READY role.data=.*
8  endmodule
```

**Figure 9: Interface `pragmas` in Verilog mapping ports with the `m_axi_` prefix to handshake interfaces and bundle ports with the same prefix (e.g., `m_axi_AW`). Suffixes `VALID` and `READY` indicate port roles, while any other suffixes denote data.**

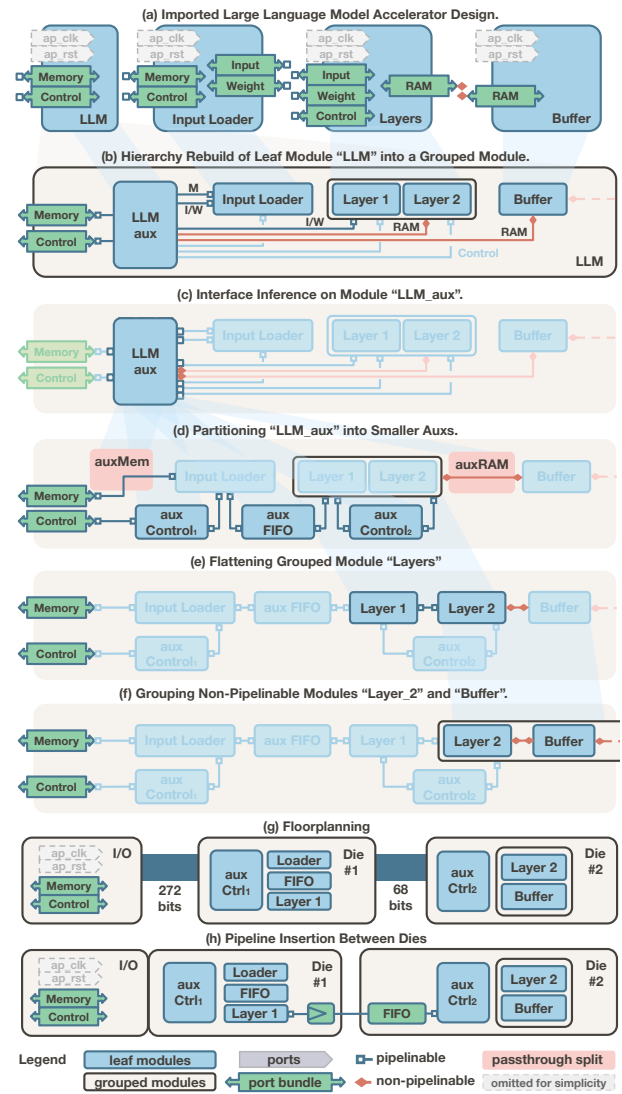

**Figure 10: RIR passes on the LLM accelerator example.**

module ports as the original leaf module, all of which are connected to the aux module. At this stage, the pass does *not* analyze the interconnections between submodules, but instead adds corresponding ports on the aux module for each port of the original submodules.

Figure 10b shows the rebuild pass applied to LLM, restructuring it into a grouped module containing its submodules and an aux module, LLM_Aux. This aux module contains the control logic and interconnects of LLM. Directly analyzing LLM's interconnect is challenging due to the complexity of its source format, which includes Verilog language syntax such as always and generate, requiring a full elaborator. Maintaining and updating such an elaborator for various design formats would be labor-intensive.

This transformation pass preserves the IR assumptions (§3.1):

(1) In the newly formed grouped module, each wire is connected to exactly two modules: the aux module and a submodule.
(2) (a) Ports of an extracted submodule are connected to a wire identifier that interconnects with the aux module.
    (b) Each port of the aux module is directly connected by a wire to an extracted submodule or to a port on the restructured grouped module; in either case, it is an identifier.
(3) Every port on a submodule is wholly connected to the aux module, ensuring that there is no splitting of the interface.

The implementation is straightforward for *any source format* if there exists a rewriter providing three functionalities: (1) extraction of submodule names and port connections; (2) addition of new ports to a module; and (3) connection of expressions to these new ports.

For example, the Slang [35] tool allows for the extraction of Verilog submodule information; new ports are added by modifying the syntax tree; connections are rerouted by appending new assign statements. This approach is adaptable to other source formats. Note that even without a dedicated rewriter for a particular source format, RIR can still manage modules in this format by treating them as leaf modules; thus, it is still possible to insert pipeline stages between these modules or to partition them as needed.

**Interface Inference Pass.** When design modules lack explicit interface information necessary for pipeline insertion, this pass infers interfaces from other modules. For instance, a user might have a grouped module where all ports are directly connected to submodules, yet the module itself lacks interface data. By leveraging the interface details from these submodules, the interface inference pass can deduce the interface for the parent module.

Interface information propagates not only between parent and child modules but also among siblings. Specifically, for aux modules created during the hierarchy rebuild pass, the interface inferencer defines their interfaces by transferring information from the aux's sibling modules, the extracted submodules, thus completing the aux module's interface, as shown in Figure 10c.

**Partitioning Pass.** This pass divides a leaf module into splits for separate floorplanning. It effectively serves as a communication analysis pass in conjunction with the hierarchy rebuild pass. After the rebuild pass, it splits the aux module created by the rebuild pass to decentralize submodule communications. Figure 10d shows the partitioning of the LLM_aux module, which initially connects all submodules. After partitioning, it is divided into five splits, including memory connections (auxMem and auxRAM) and control logic (auxControl$_1$ and auxControl$_2$). In this example, the LLM module implements FIFO logic in the Verilog body, connecting the input loader and the first layer, which is split into auxFIFO.

It converts modules in *arbitrary formats* to netlists using EDA flows and applies union-find [15] written with RapidWright [27] to analyze port connectivity, excluding clock and reset signals due to their shared use in submodules. Disjoint components are separated into new splits. The splits are created by *wrapping* the original aux module, exposing only the necessary ports and preserving the internal logic. Unconnected logic remains undriven, which will be eliminated by subsequent EDA flows. The new splits replace the original aux in the IR, and clock and reset signals are distributed to all submodules through dedicated broadcasting aux modules.

It maintains IR assumptions by merging ports in a common interface into a union-set, preventing the interface from spanning multiple splits. It introduces no new connections except for clock and reset signals, which are managed by broadcasting modules. This ensures that, post-transformation, all wires still connect exactly two modules, and all ports connect to either an identifier or a constant.

**Passthrough Pass.** If netlist analysis shows that an interface connects solely and directly to another, the module can be bypassed by rerouting connections between interfaces. In Figure 10d, the auxRAM split is bypassed, allowing direct connections between the Layer_2 and Buffer modules. This simplifies the IR and reduces the number of modules, making the design more readable and easier to optimize. The passthrough pass maintains IR assumptions by detaching a wire from one module before connecting it to another.

**Flattening Pass.** HLPS optimization formulations, such as integer linear programming (ILP) used in AutoBridge [17], often require a flat graph rather than a hypergraph with multiple hierarchical levels. The flattening pass transforms a hierarchical design into a flat one by recursively merging all grouped modules into a single one. During this process, wires are consolidated, and submodules and their connections are reestablished in the new module. Figure 10e shows the flattening pass incorporating the Layer_1 and Layer_2 submodules into the LLM module. Without this pass, these two modules would have to be grouped into a single partition, resulting in suboptimal partitioning since they are both resource-intensive.

This pass adheres to the IR assumptions by not introducing new interconnections between modules. It solely consolidates existing wires and submodules, thus preserving the original properties.

**Wrapping Pass.** This pass uses a template to wrap a module. Within the template, helper submodules can be added alongside the wrapped module. The wrapper ports can be connected to either the helpers or the wrapped module. This pass can implement partitioning by exposing only specific ports. It can also add pipeline stages as helper submodules. Typically, a flattening pass follows to elevate the helpers, effectively *inserting* the helper modules.

**Grouping Pass.** This pass restructures a flat design into a hierarchy. In Figure 10f, submodules Layer_2 and Buffer are grouped into a new grouped module. This pass can be used to merge non-pipelinable modules and to specify floorplanning constraints.

### 3.4 Framework Integration

We develop a complete HLPS system in RIR following the methodology described in Section 2.2, integrating our plugins and passes to assess RIR's applicability. Figure 10 illustrates the tool's workflow, which consists of the four stages of the HLPS methodology.

(1) **Communication Analysis**: The tool (a) imports design and interface data into RIR, (b) restructures large modules into grouped modules using the hierarchy rebuild pass, (c) infers interfaces for aux modules and for modules lacking this information, and (d) partitions broadcasting modules and applies passthrough, especially for aux modules. This stage captures coarse-grained communication patterns between modules.

(2) **Design Partitioning**: The tool (e) converts the design into a flat representation and (f) groups non-pipelined modules with adjacent ones. This stage partitions the design into pipelinable sections according to the identified communication patterns.

(3) **Coarse-Grained Floorplanning**: Utilizing AutoBridge's integer linear programming (ILP) formulation [17], the tool (g) optimizes the placement of modules into predefined slots on a virtual device and designs a pipeline insertion scheme. This stage allocates partitions to coarse regions on the FPGA, aiming to minimize cross-region wiring and adhere to constraints such as DSP count and the number of boundary-crossing wires.

(4) **Global Interconnect Synthesis**: Following floorplanning, the tool clusters modules in the same region using the grouping pass. It then (h) inserts pipeline stages with the wrapping pass. This stage generates inter-partition connections based on estimated delay to break critical paths and aid in timing closure. Finally, the optimized design is exported for implementation.

In summary, we incorporated AutoBridge's formulation [17] into our RapidStream IR infrastructure, enhancing it as a flexible and modular physical synthesis tool for new exploration strategies and the composition of different source formats and devices. Section 4 demonstrates the framework's adaptability and assesses frequency enhancements in FPGA designs and devices, comparing outcomes with original vendor tools and manual optimizations.

## 4 EVALUATION

We evaluated RapidStream IR (RIR) targeting AMD FPGAs using Vivado 2023.2. Experiments were conducted on an AMD EPYC 7282 CPU, 128 GB of RAM, and Ubuntu 22.04. We utilized the COIN-OR solver [36] with a 400-second limit for ILP optimization tasks. For diverse design format tests, Dynamatic 2.0, Catapult HLS 2021.1, and Intel FPGA HLS 19.4.0 were used to produce RTL inputs. We seek to answer the following research questions:

**RQ1** Can RIR passes be used to transform FPGA designs in various input formats, including handcrafted Verilog and HLS designs generated by different vendor tools?

**RQ2** Does RIR effectively reduce the effort required for programmers to implement new research exploration tasks?

**RQ3** Is RIR capable of providing frequency improvements for complex FPGA designs and new target devices?

### 4.1 Custom HLS Input

Dynamatic is an open-source compiler that converts C++ code into VHDL designs with dynamic scheduling using handshake protocols [22, 23]. Commercial tools such as Catapult HLS [37] and Intel HLS [21] also create handshakes via custom data types. Prior work lacks the infrastructure to manipulate the generated RTL designs. To develop a new frontend in RIR that accepts the generated designs, three components are needed: (1) a metadata parser, (2) an interface analyzer, and (3) a code rewriter.

```
1  add_reset(module=".*", port="rst|reset", active="high")
2  add_handshake(module=top_level, pattern="{bundle}_{role}",
3      role={ready:"ready", valid:"valid", data:"in|out"})
```

**Figure 11: Snippet of the interface rules for Dynamatic.**

**Table 1: Code in Python or Verilog for supporting HLS tools.**

| Software      | Dynamatic | Catapult HLS | Intel HLS |
|---------------|-----------|--------------|-----------|
| **Lines of code** | 146       | 158          | 204       |

The metadata parser and code rewriter are common to most input that use standard hardware description languages such as Verilog or VHDL. The interface analyzer, however, is specific to each HLS framework. This subsection focuses on the interface analyzer.

**Dynamatic** has elastic elements with consistent naming, aligning well with the *interface rules* (§3.2). We use 20 rules in Python to specify *all* its handshake interfaces. Figure 11 shows two of them: one specifies reset signals using the regular expression ".*" to match and apply to all modules, and the other defines the handshakes of the top-level module. **Catapult HLS** synthesizes handshakes using customizable design libraries such as ccs_out_wait and ccs_in_wait; with simple pragmas in these modules' Verilog code, the interface can be automatically propagated during the interface inference pass to neighboring modules. **Intel HLS** creates handshakes mostly with consistent port naming, making them also compatible with the Python-based interface rules method.

The total lines of Python or Verilog code required for RIR to handle inputs from these HLS tools are shown in Table 1. We conducted experiments with benchmarks from three sources: all 29 examples from the Dynamatic repository [14], a sparse linear algebra accelerator for Catapult HLS [13], and all 12 benchmarks from the CHStone suite for Intel HLS [11]. Our approach successfully extracted the interface information. Additionally, our parser and code rewriter effectively imported designs from all benchmarks into the RIR, transforming their hierarchy, inserting pipelines, and exporting a functionally equivalent RTL design.

> **Summary 1 (RQ1: Input Formats)**
>
> By extending RIR to support RTL from various HLS tools, we demonstrate its ability to handle multiple input formats.

### 4.2 Floorplan Exploration

Floorplanning requires tradeoffs between local and global optimization targets, as shown in Figure 12, which lists ten different floorplans for the example LLM design [8]. The line chart in the

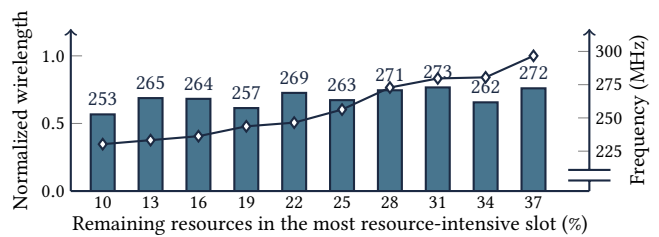

**Figure 12: Relationship between resource distribution, wirelength, and frequency for the LLM design on VHK158.**

Table 2: Frequency improvements automated with RapidStream IR for various design formats on different FPGAs.

| Application | Target | Benchmark Features | | | LUT (%) | FF (%) | BRAM (%) | DSP (%) | URAM (%) | Freq (MHz) | | |
|---|---|---|---|---|---|---|---|---|---|---|---|---|
| | | Hierarchy | Mixed-Source | New FPGAs | | | | | | Original | RIR | Others |
| CNN 13×4 | U250 | | | | 13 | 11 | 10 | 17 | 0 | 233 | 335 (+44%) | 325 [17] |
| CNN 13×6 | U250 | | | | 15 | 16 | 13 | 26 | 0 | 234 | 327 (+40%) | 324 [17] |
| CNN 13×8 | U250 | | | | 26 | 22 | 16 | 24 | 0 | 245 | 332 (+36%) | 320 [17] |
| CNN 13×10 | U250 | | | | 30 | 27 | 28 | 43 | 0 | - | 320 (+∞%) | 322 [17] |
| CNN 13×12 | U250 | | | | 27 | 33 | 30 | 51 | 0 | - | 305 (+∞%) | 295 [17] |
| LLaMA2 | VP1552 | ✓ | ✓ | ✓ | 32 | 16 | 13 | 22 | 18 | 198 | 258 (+30%) | N/A |
| LLaMA2 | VHK158 | ✓ | ✓ | ✓ | 32 | 16 | 13 | 22 | 18 | 206 | 273 (+33%) | N/A |
| LLaMA2 | U55C | ✓ | ✓ | ✓ | 49 | 25 | 24 | 18 | 24 | 165 | 247 (+50%) | N/A |
| LLaMA2 | VU9P | ✓ | ✓ | | 59 | 32 | 23 | 24 | 24 | 141 | 212 (+50%) | N/A |
| LLaMA2 | U250 | ✓ | ✓ | | 42 | 23 | 20 | 14 | 19 | 159 | 228 (+43%) | N/A |
| LLaMA2 | U280 | ✓ | ✓ | | 49 | 25 | 24 | 18 | 25 | 150 | 243 (+62%) | 245 [8] |
| LLaMA2 (opt) | U280 | ✓ | ✓ | | 35 | 19 | 15 | 18 | 25 | 201 | 306 (+52%) | 245 [8] |
| Minimap2 | VP1552 | ✓ | | ✓ | 39 | 15 | 10 | 31 | 0 | 265 | 285 (+8%) | N/A |
| KNN | U280 | | ✓ | | 56 | 28 | 10 | 14 | 0 | - | 292 (+∞%) | N/A |
| Average Treating Unroutable Designs as Zeros | | | | | 36 | 22 | 18 | 24 | 11 | 157 | 283 (+80%) | |
| Average Excluding Originally Unroutable Designs | | | | | 36 | 20 | 16 | 21 | 14 | 200 | 277 (+39%) | |

figure shows that decreasing the amount of logic in the most congested area of the floorplan reduces local congestion but potentially leads to longer wire lengths, which adversely affect global routing results, and vice versa. Additionally, the bar chart in the same figure highlights the complexity of these tradeoffs, indicating a variation in the operating frequency of up to 20 MHz depending on the chosen tradeoff point between local and global optimization.

Without RIR, designers would need to manually explore the design space by partitioning the design, restructuring the hierarchy as previously shown in Figure 1, modifying the floorplan constraints, and re-executing the synthesis and place-and-route processes.

As an evaluation of RIR's applicability, we applied our methodology to this floorplan exploration task. By adjusting the maximum allowable resource utilization for each slot using the virtual device model described in Section 3.1, RIR optimizes the wire length in placements given the constraints. In this way, RIR automatically explores the design space of trade-offs and approximates Pareto optimality. This approach creates a variety of floorplans, as we have presented in Figure 12, allowing designers to evaluate the balance between wire length and resource distribution. This automation is implemented as a standalone RIR plugin, written in 207 lines of Python code, that can be reused across different designs. In contrast, manual exploration of this design alone would require a significant rewrite of the RTL code, consisting of hundreds of lines, potentially introducing errors and requiring numerous iterations.

> **Summary 2 (RQ2: Extensions)**
> RIR simplifies the extension of high-level physical optimizations, such as the exploration of different floorplan schemes.

### 4.3 Parallel Synthesis

In Section 3.4, we divide the design into several coarse-grained groups, each corresponding to a device slot. This approach intrinsically spawns the potential to perform parallel synthesis, where slots can be synthesized in parallel. The top-level module can be synthesized along with these slots by marking the slots as black boxes. Finally, we assemble these post-synthesis netlists to obtain the complete design. We implement the parallel synthesis program as a standalone RIR backend plugin in 299 lines of Python.

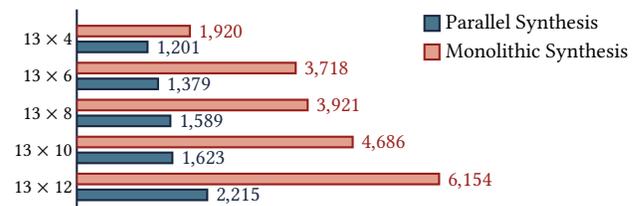

Figure 13: Synthesis wall time in seconds.

We evaluate the RIR parallel synthesis plugin using HLS benchmarks of systolic array architectures for convolutional neural networks, which are generated using AutoSA [40]. The evaluation is performed on the Alveo U250 FPGA by synthesizing the device slots in parallel, as shown in Figure 13. For systolic processing element arrays with sizes ranging from 13 × 4 to 13 × 12, the plugin achieves an average synthesis wall time acceleration of 2.49×.

> **Summary 3 (RQ2: Extensions)**
> RIR allows easy design transformation, opening up potential for EDA tool research with the divide-and-conquer methodology.

### 4.4 Benchmarking

We evaluate our HLPS tool, developed in RIR, on a set of real-world FPGA designs that fail to meet timing targets. The frequency results are compared with those obtained from AutoBridge [17] and the standard EDA tool supplied by the FPGA vendor, AMD Vivado.

(1) **CNN** refers to a convolutional neural network accelerator designed with AutoSA [40] into a systolic array architecture in Vitis HLS. It features a flat hierarchy, which is supported by AutoBridge. We use this benchmark to compare the frequency results between RIR and AutoBridge.

(2) **LLaMA2** refers to a hybrid-source accelerator designed for large language model inference of the LLaMA2 model, initially optimized for the Alveo U280 FPGA with a four-level nested pipeline using HLS, Xilinx IPs, and manual RTL [8, 9]. AutoBridge does not support it due to its complex hierarchical design. We further ported it to Versal boards using RIR and compared the frequency performance with that of Vivado, demonstrating RIR's adaptability for multi-source and multi-target designs.

(3) **Minimap2** refers to an accelerator for long-read genome sequencing with multiple hierarchical levels of pipelines, initially developed for UltraScale+ VU9P using Vitis HLS [19]. In the benchmarking, we retained the original hierarchical structure of Minimap2 and ported it to the AMD Versal VP1552 device.
(4) **KNN** refers to a k-nearest neighbor accelerator for the Alveo U280 FPGA, using HLS kernels and a custom RTL interconnect, implemented on the Vitis platform [29]. RIR directly ingests the Vitis-packed Xilinx Object (XO) files for optimization and outputs the optimized design in the same format, acting as a transparent plugin to the Vitis framework.

RIR successfully ingests all designs and applies transformations using the HLPS methodology, as summarized in Table 2. The design features, such as multiple levels of pipelined hierarchy (Hierarchy), a mixture of different source formats (Mixed-Source), including RTL, HLS, and IP, and implementation targets for new FPGA devices (New FPGAs), are listed in the "Benchmark Features" columns. These features are unsupported by existing HLPS frameworks. In the "Freq (MHz)" columns, the "Original" column shows the frequency of the original design before optimizations implemented using Vivado; the "RIR" column presents the frequency of designs optimized by RIR; and the "Other" column includes results from existing literature [8, 17]. We indicate the frequency results for benchmarks that fail routing as "-". For FPGA resources, we report the original utilization percentages on the target device. The change in resource post-optimization is within 1% across all benchmarks.

> **Summary 4 (RQ1: Input Formats)**
>
> RIR efficiently rewrites FPGA designs from various formats, such as handcrafted RTL, IP-integrated projects, and sophisticated designs with multiple levels of pipelinable hierarchy.

Table 2 also compares our work with AutoBridge [17] and manual optimization [8] in the last column. Our HLPS transformations achieve frequency improvements similar to AutoBridge on "*CNN*" designs for Alveo U250 and 30% to 62% improvements on porting "*LLaMA2*" across various devices. Our "*LLaMA2*" frequency on U280 is comparable to manual optimization. Further refactoring "*LLaMA2*" into "*LLaMA2 (opt)*" by decomposing HLS functions into smaller pipelinable parts boosts the frequency to 306 MHz.

> **Summary 5 (RQ3: Frequency Improvements)**
>
> RIR has demonstrated cutting-edge frequency improvements compared to state-of-the-art solutions, while supporting a much broader range of input formats and target FPGA devices.

## 5 RELATED WORK

A significant body of literature has explored HLPS methodologies [12, 16, 17, 25, 30, 32, 33, 42, 43]. AutoBridge [17] improves timing by considering layout information during HLS stages, creating high-frequency multi-die FPGA HLS designs. However, it supports only pipelining with streaming ports at the top-level function in a dataflow manner. In contrast, RapidStream IR (RIR) supports pipelining at arbitrary hierarchy levels in hybrid-source designs. Section 4 demonstrates that AutoBridge's methodologies can be integrated into our framework as a pass, enhancing its functionality with our multi-level pipeline parsing without performance loss.

Other existing HLPS works can be integrated into our framework, providing them with broader support for input and target devices.

Despite the significant progress in HLPS, a reusable HLPS IR is lacking. Existing IRs or languages, such as MLIR [26], CIRCT [28], Yosys IR [41], ScaleHLS [44], Chisel [4], Calyx [34], CIRRF [39], HIR [31], Allo [9], and Xilinx IPI [2], address various aspects of compilation, circuit logic, HLS, datapaths, schedules, and IP integration. Nevertheless, they do not provide the necessary infrastructure to support existing designs and pipeline coarse-grained partitions, which are unique challenges in HLPS. These frameworks are orthogonal to RIR and can be integrated into RIR as leaf modules, allowing us to reap the benefits of both worlds. Section 3.1 provides additional discussion on this comparison.

## 6 DISCUSSION AND CONCLUSION

In this paper, we introduce RapidStream IR (RIR), which is accessible to academic researchers for free on https://rapidstream-da.com. All case studies discussed in Section 4 are implemented as standalone plugins or Python scripts without modifying the core infrastructure. We anticipate a wide range of use cases for RIR and propose several potential future research directions:

(1) **Automated NoC Synthesis:** RIR could enable NoC synthesis for FPGA HLS designs by automatically integrating routers between handshake modules. Challenges include analyzing intra-node patterns for bandwidth needs, managing traffic, and optimizing the network to minimize the impact on throughput.
(2) **Parallel Placement and Routing:** RapidStream [20] enables parallel placement and routing for flat dataflow FPGA HLS designs with Vitis HLS. RIR provides hierarchy flattening and reorganization, accommodating multiple hierarchical levels and various HLS tools, which could enhance RapidStream. Challenges include interfacing with vendor tools and developing a custom placer and router using RapidWright [27].
(3) **Design Instrumentation:** RIR could be extended to automate the insertion of performance counters and monitoring IPs, placed between modules using interface information. This would help in on-board design profiling, pinpointing performance bottlenecks, and analyzing behaviors such as bandwidth requirements for NoCs. It could also support fuzz verification by randomly throttling identified handshakes.

RIR is a versatile infrastructure for developing high-level physical synthesis (HLPS) tools for FPGAs. It features a progressively refined intermediate representation, supports plugins for vendor tool integration, and enables reusable transformation passes. RIR improves FPGA design frequency by 30% to 62% across diverse benchmarks and platforms. Additionally, some initially unroutable designs are able to achieve frequencies around 300 MHz after RIR's HLPS optimizations. RIR is extensible for research, potentially fostering innovative future studies in the FPGA EDA community.

## ACKNOWLEDGMENTS

We thank Chris Lavin, Eddie Hung, Pongstorn Maidee, DJ Wang and Luciano Lavagno from AMD for their technical insights, and Hongzheng Chen for his help with the LLM accelerator and the Dynamatic HLS team with benchmarks and code generation. J. Cong and Z. Zhiru serve on the technical advisory board at RapidStream.

# REFERENCES

[1] Advanced Micro Devices, Inc. 2024. *Versal Adaptive SoC Technical Reference Manual.* https://docs.amd.com/r/en-US/am011-versal-acap-trm
[2] Advanced Micro Devices, Inc. 2024. *Vivado Design Suite User Guide: Designing with IP.* https://docs.amd.com/r/en-US/ug896-vivado-ip/IP-Integrator
[3] Advanced Micro Devices, Inc. 2024. *Xilinx Intellectual Property.* https://www.xilinx.com/products/intellectual-property.html
[4] Jonathan Bachrach, Huy Vo, Brian Richards, Yunsup Lee, Andrew Waterman, Rimas Avižienis, John Wawrzynek, and Krste Asanović. 2012. Chisel: constructing hardware in a Scala embedded language. In *Proceedings of the 49th Annual Design Automation Conference (DAC '12).*
[5] Oren Ben-Kiki, Clark Evans, and Brian Ingerson. 2009. Yaml ain't markup language (yaml™) version 1.1. *Working Draft 2008* 5, 11 (2009).
[6] Julien Boucaron, Anthony Coadou, and Robert de Simone. 2009. Latency-Insensitive Design: Retry Relay-Station and Fusion Shell. In *Proceedings of the 4th International Workshop on the Application of Formal Methods for Globally Asynchronous and Locally Synchronous Design (FMGALS '09)*, Vol. 245. 23–33.
[7] Pierre Bourhis, Juan L. Reutter, Fernando Suárez, and Domagoj Vrgoč. 2017. JSON: Data model, Query languages and Schema specification *(PODS '17).*
[8] Hongzheng Chen, Jiahao Zhang, Yixiao Du, Shaojie Xiang, Zichao Yue, Niansong Zhang, Yaohui Cai, and Zhiru Zhang. 2024. Understanding the Potential of FPGA-Based Spatial Acceleration for Large Language Model Inference. *ACM Transactions on Reconfigurable Technology and Systems* (Apr 2024).
[9] Hongzheng Chen, Niansong Zhang, Shaojie Xiang, Zhichen Zeng, Mengjia Dai, and Zhiru Zhang. 2024. Allo: A Programming Model for Composable Accelerator Design. In *Proceedings of the ACM on Programming Languages (PLDI '24).*
[10] Jason Cong, Jason Lau, Gai Liu, Stephen Neuendorffer, Peichen Pan, Kees Vissers, and Zhiru Zhang. 2022. FPGA HLS Today: Successes, Challenges, and Opportunities. *ACM Transactions on Reconfigurable Technology and Systems* 15, 4, Article 51 (Aug 2022), 42 pages.
[11] Steve Dai, Gai Liu, and Zhiru Zhang. 2018. A Scalable Approach to Exact Resource-Constrained Scheduling Based on a Joint SDC and SAT Formulation. In *Proceedings of the 2018 ACM/SIGDA International Symposium on Field-Programmable Gate Arrays (FPGA '18).*
[12] Linfeng Du, Tingyuan Liang, Sharad Sinha, Zhiyao Xie, and Wei Zhang. 2023. FADO: Floorplan-Aware Directive Optimization for High-Level Synthesis Designs on Multi-Die FPGAs. In *Proceedings of the 2023 ACM/SIGDA International Symposium on Field Programmable Gate Arrays (FPGA '23).*
[13] Yixiao Du. 2024. Cornell University: Building Sparse Linear Algebra Accelerators with HLS. Webinar. https://webinars.sw.siemens.com/en-US/cornell-university-building-sparse-linear-algebra-accelerators-with-hls/
[14] EPFL Processor Architecture Laboratory. 2024. DHLS (Dynamic High-Level Synthesis) Compiler Based on MLIR. https://github.com/EPFL-LAP/dynamatic/
[15] Bernard A. Galler and Michael J. Fisher. 1964. An improved equivalence algorithm. *Commun. ACM* 7, 5 (May 1964), 301–303.
[16] Licheng Guo, Yuze Chi, Jason Lau, Linghao Song, Xingyu Tian, Moazin Khatti, Weikang Qiao, Jie Wang, Ecenur Ustun, Zhenman Fang, Zhiru Zhang, and Jason Cong. 2023. TAPA: A Scalable Task-parallel Dataflow Programming Framework for Modern FPGAs with Co-optimization of HLS and Physical Design. *ACM Transactions on Reconfigurable Technology and Systems*, Article 63 (Dec 2023).
[17] Licheng Guo, Yuze Chi, Jie Wang, Jason Lau, Weikang Qiao, Ecenur Ustun, Zhiru Zhang, and Jason Cong. 2021. AutoBridge: Coupling Coarse-Grained Floorplanning and Pipelining for High-Frequency HLS Design on Multi-Die FPGAs. In *Proceedings of the 2021 ACM/SIGDA International Symposium on Field-Programmable Gate Arrays (FPGA '21).*
[18] Licheng Guo, Jason Lau, Yuze Chi, Jie Wang, Cody Hao Yu, Zhe Chen, Zhiru Zhang, and Jason Cong. 2020. Analysis and Optimization of the Implicit Broadcasts in FPGA HLS to Improve Maximum Frequency. In *Proceedings of the 2020 57th ACM/IEEE Design Automation Conference (DAC '20).*
[19] Licheng Guo, Jason Lau, Zhenyuan Ruan, Peng Wei, and Jason Cong. 2019. Hardware Acceleration of Long Read Pairwise Overlapping in Genome Sequencing: A Race Between FPGA and GPU. In *Proceedings of the IEEE 27th International Symposium on Field-Programmable Custom Computing Machines (FCCM '19).*
[20] Licheng Guo, Pongstorn Maidee, Yun Zhou, Chris Lavin, Jie Wang, Yuze Chi, Weikang Qiao, Alireza Kaviani, Zhiru Zhang, and Jason Cong. 2022. RapidStream: Parallel Physical Implementation of FPGA HLS Designs. In *Proceedings of the 2022 ACM/SIGDA International Symposium on Field-Programmable Gate Arrays (FPGA '22).*
[21] Intel. 2024. Intel High Level Synthesis Compiler. https://www.intel.com/content/www/us/en/software/programmable/quartus-prime/hls-compiler.html
[22] Lana Josipović, Radhika Ghosal, and Paolo Ienne. 2018. Dynamically Scheduled High-level Synthesis. In *Proceedings of the 2018 ACM/SIGDA International Symposium on Field-Programmable Gate Arrays (FPGA '18).*
[23] Lana Josipović, Andrea Guerrieri, and Paolo Ienne. 2020. Dynamatic: From C/C++ to Dynamically Scheduled Circuits. In *Proceedings of the 2020 ACM/SIGDA International Symposium on Field-Programmable Gate Arrays (FPGA '20).*
[24] JSON Schema. 2020. JSON Schema: A Media Type for Describing JSON Documents. https://json-schema.org/specification.html. Draft 2020-12.
[25] Moazin Khatti, Xingyu Tian, Yuze Chi, Licheng Guo, Jason Cong, and Zhenman Fang. 2023. PASTA: Programming and Automation Support for Scalable Task-Parallel HLS Programs on Modern Multi-Die FPGAs. In *Proceedings of the 2023 IEEE 31st Annual International Symposium on Field-Programmable Custom Computing Machines (FCCM '23).*
[26] Chris Lattner, Mehdi Amini, Uday Bondhugula, Albert Cohen, Andy Davis, Jacques Pienaar, River Riddle, Tatiana Shpeisman, Nicolas Vasilache, and Oleksandr Zinenko. 2021. MLIR: Scaling Compiler Infrastructure for Domain Specific Computation. In *Proceedings of the 2021 IEEE/ACM International Symposium on Code Generation and Optimization (CGO '21).*
[27] Chris Lavin and Alireza Kaviani. 2018. RapidWright: Enabling custom crafted implementations for FPGAs. In *Proceedings of the IEEE 26th Annual International Symposium on Field-Programmable Custom Computing Machines (FCCM '18).*
[28] LLVM. [n. d.]. CIRCT: Circuit IR Compilers and Tools. https://circt.llvm.org/.
[29] Alec Lu, Zhenman Fang, Nazanin Farahpour, and Lesley Shannon. 2020. CHIP-KNN: A Configurable and High-Performance K-Nearest Neighbors Accelerator on Cloud FPGAs. In *Proceedings of the 2020 International Conference on Field-Programmable Technology (ICFPT '20).*
[30] Jianwen Luo, Xinzhe Liu, Fupeng Chen, and Yajun Ha. 2023. HRFF: Hierarchical and Recursive Floorplanning Framework for NoC-Based Scalable Multidie FPGAs. *IEEE Transactions on Circuits and Systems I: Regular Papers* 70, 11 (2023).
[31] Kingshuk Majumder and Uday Bondhugula. 2024. HIR: An MLIR-based Intermediate Representation for Hardware Accelerator Description. In *Proceedings of the 28th ACM International Conference on Architectural Support for Programming Languages and Operating Systems (ASPLOS '23).*
[32] Mohammadmahdi Mazraeli, Yu Gao, and Paul Chow. 2023. Partitioning Large-Scale, Multi-FPGA Applications for the Data Center. In *Proceedings of the 33rd International Conference on Field-Programmable Logic and Applications (FPL '23).*
[33] Tan Nguyen, Zachary Blair, Stephen Neuendorffer, and John Wawrzynek. 2023. SPADES: A Productive Design Flow for Versal Programmable Logic. In *Proceedings of the 33rd Conference on Field-Programmable Logic and Applications (FPL '23).*
[34] Rachit Nigam, Samuel Thomas, Zhijing Li, and Adrian Sampson. 2021. A compiler infrastructure for accelerator generators. In *Proceedings of the 26th ACM International Conference on Architectural Support for Programming Languages and Operating Systems (ASPLOS '21).*
[35] Michael Popoloski. 2024. Slang: SystemVerilog compiler and language services. https://github.com/MikePopoloski/slang.
[36] Matthew J. Saltzman. 2002. *Coin-OR: An Open-Source Library for Optimization.* Springer US, Boston, MA, 3–32.
[37] Siemens. 2024. *Catapult High-Level Synthesis and Verification.* https://static.sw.cdn.siemens.com/siemens-disw-assets/public/2viQ3qHCWJQxzqSkwBauMQ/en-US/Siemens-SW-Catapult-HLS-HLV-Platform-FS-82981-D1.pdf
[38] Jérôme Siméon and Philip Wadler. 2003. The essence of XML. *SIGPLAN Not.* 38, 1 (January 2003), 1–13.
[39] Jason Villarreal, Adrian Park, Walid Najjar, and Robert Halstead. 2010. Designing Modular Hardware Accelerators in C with ROCCC 2.0. In *Proceedings of the 2010 18th IEEE Annual International Symposium on Field-Programmable Custom Computing Machines (FCCM '10).*
[40] Jie Wang, Licheng Guo, and Jason Cong. 2021. AutoSA: A Polyhedral Compiler for High-Performance Systolic Arrays on FPGA. In *Proceedings of the 2021 ACM/SIGDA International Symposium on Field-Programmable Gate Arrays (FPGA '21).*
[41] Clifford Wolf, Johann Glaser, and Johannes Kepler. 2013. Yosys: A free Verilog synthesis suite. In *Proceedings of the 21st Austrian Workshop on Microelectronics (Austrochip).*
[42] Yuanlong Xiao, Aditya Hota, Dongjoon Park, and André DeHon. 2022. HiPR: High-level Partial Reconfiguration for Fast Incremental FPGA Compilation. In *Proceedings of the 2022 32nd International Conference on Field-Programmable Logic and Applications (FPL '22).*
[43] Yuanlong Xiao, Dongjoon Park, Zeyu Jason Niu, Aditya Hota, and André Dehon. 2024. ExHiPR: Extended High-Level Partial Reconfiguration for Fast Incremental FPGA Compilation. *ACM Transactions on Reconfigurable Technology and Systems* 17, 2, Article 21 (Mar 2024), 28 pages.
[44] Hanchen Ye, Cong Hao, Jianyi Cheng, Hyunmin Jeong, Jack Huang, Stephen Neuendorffer, and Deming Chen. 2022. ScaleHLS: A New Scalable High-Level Synthesis Framework on Multi-Level Intermediate Representation. In *Proceedings of the 2022 IEEE International Symposium on High-Performance Computer Architecture (HPCA '22).*
[45] Shulin Zeng, Jun Liu, Guohao Dai, Xinhao Yang, Tianyu Fu, Hongyi Wang, Wenheng Ma, Hanbo Sun, Shiyao Li, Zixiao Huang, Yadong Dai, Jintao Li, Zehao Wang, Ruoyu Zhang, Kairui Wen, Xuefei Ning, and Yu Wang. 2024. FlightLLM: Efficient Large Language Model Inference with a Complete Mapping Flow on FPGAs. In *Proceedings of the 2024 ACM/SIGDA International Symposium on Field Programmable Gate Arrays (FPGA '24).*